\documentclass[aps,pra]{revtex4}
\usepackage{graphicx}

\begin{document}

\title{Minimal classical communication and measurement complexity
for quantum information splitting}
\author{Zhan-jun Zhang$^{a,b,*}$, Chi-Yee Cheung$^{c,\dag}$\\
{\normalsize $^a$ Department of Physics and Center for
Quantum Information Science,}\\
{\normalsize National Cheng Kung University, Tainan 70101, Taiwan}\\
{\normalsize $^b$ Key Laboratory of Optoelectronic Information
Acquisition \& Manipulation of Ministry of Education of China,}\\
{\normalsize School of Physics \& Material Science, Anhui
University, Hefei 230039, China} \\{\normalsize $^c$ Institute of
Physics, Academia Sinica, Taipei 11529, Taiwan} \\{\normalsize
$^{*}$zjzhang@ahu.edu.cn}~~ $^{\dag}$cheung@phys.sinica.edu.tw}

\maketitle

\begin{minipage}{420pt}

{\bf Abstract} We present two quantum information splitting schemes
using respectively tripartite GHZ and asymmetric W states as quantum
channels. We show that, if the secret state is chosen from a special
ensemble and known to the sender (Alice), then she can split and
distribute it to the receivers Bob and Charlie by performing only a
single-qubit measurement and broadcasting an one-cbit message. It is
clear that no other schemes could possibly achieve the same goal
with simpler measurement and less classical communication. In
comparison, existing schemes work for arbitrary quantum states which
need not be known to Alice, however she is required to perform a
two-qubit Bell measurement and communicate a two-cbit message. Hence
there is a trade off between flexibility and measurement complexity
plus classical resource. In situations where our schemes are
applicable, they will greatly reduce the measurement complexity and
at the same time cut the communication overhead by one half.\\



\end{minipage}

\noindent {\bf 1 Introduction}

Understanding the minimal amount of resources required to implement
a task is a fundamental issue in quantum information theory. It has
been emphasized that, in addition to quantum resources, it is
equally important to take into account the role of classical
communication involved[1-3]. A prime example is provided by the
novel scheme of quantum teleportation introduced by Bennett et
al.[4] in 1993, which has attracted much attention over the
years[5-15]. In this scheme, with the aid of an
Einstein-Podolsky-Rosen (EPR) pair shared between two remote sites,
an arbitrary unknown quantum state can be teleported from one site
to the other without physically transmitting any particles between
them. All the sender needs to do is to perform a Bell-state
measurement and publicly announce the outcome which is a message
consisting of two classical bits (cbits). Upon receiving the
sender's message, the receiver can reconstruct the unknown state by
executing an appropriate unitary operation on his share of the EPR
pair. Quantum teleportation demonstrates the interchangeability of
different types of resources[5]. Mostly importantly it shows that
transmitting the information contained in one qubit consumes one
entangled bit (ebit) plus two cbits from the sender to the receiver.

In 2000, Lo[1] considered a different but related task called
"remote state preparation (RSP)" in which a sender helps a receiver
in a remote site to prepare a state chosen from a given ensemble.
Such a task is clearly useful in distributed quantum information
processing (QIP). The author studied the amount of classical
communication cost needed and found that, for certain special
ensembles of states, RSP requires less amount of classical
communication than teleportation. But he conjectured that for
general states the classical communication costs of the two tasks
would be equal. At about the same time, Pati[16] considered the
minimum number of classical bits needed for RSP and also the
complexity of the measurement involved. In 2001, Bennett et al.[17]
proved that, to remotely prepare a large number ($n$) of qubit
states, the asymptotic classical communication cost is one cbit per
qubit, which is only half that needed in teleportation. This cost
can further be reduced when the goal is to transmit only part of a
known entangled state. However, there exists no faithful RSP scheme
for finite $n$ that always uses less classical communication than
teleportation.  The protocol of RSP has continued to attract
attention from researchers in recent years [18-27].

As we saw, in RSP as well as teleportation, a prior
established quantum channel of one ebit linking the two
parties is mandatory. However, there is a key difference.
Whereas by teleportation the sender can prepare an
arbitrary unknown state in a remote site, the state to be
prepared must be known to the sender in RSP. Due to this
restriction, in RSP the sender only needs to perform a
single-qubit measurement and send a one-cbit message to the
receiver.  In contrast the sender must perform a two-qubit
Bell measurement and sends two cbits to the receiver in
teleportation.

Quantum information splitting (QIS) or quantum state sharing is the
quantum generalization of classical secret sharing. The first QIS
protocols were proposed by Hillery et al.[28] and Karlsson et al.
[29] in 1999. Since then many other protocols have been proposed,
and the topic still attracts much attention today[30-41]. In QIS, a
piece of quantum information (in the form of a quantum state) is
divided and distributed to a number of receivers. Although it is
possible to distribute quantum information by sending the qubits
directly, for security reason and also to avoid decoherence effects
in the physical channel, it is more desirable to do so via
pre-established entanglement between the sender and the receivers.
Actually in almost all existing schemes, the qubits are sent to
individual receivers via teleportation.

We note that in all existing QIS schemes, the state to be
shared need not be known to the sender. This is the reason
why teleportation is used in the distribution process.
However this may be a waste of resources in general. It is
reasonable to assume that, in most situations, the sender
in a QIS protocol is the the owner of the secret
information being distributed; hence he/she knows the
secret. If this is the case, one could use a RSP (instead
of teleportation) scheme to distribute the secret qubits.
Clearly this will save the amount of classical
communication needed, and at the same time reduced the
complexity of the quantum measurement involved.

In this spirit, we consider in this paper two QIS schemes
in which the sender knows exactly what the the secret
quantum information is. Specifically we study tripartite
cases where the three legitimate parties share a GHZ or
asymmetry W state. The secret quantum information is
assumed to be chosen from an ensemble of states located on
the equatorial or polar great circle on the Bloch sphere.
They will be called ``equatorial" and ``real" states
respectively in the following. Incidentally, the idea in
this paper is important for it can be also applied to
improve other QIS schemes to reduce the measurement
complexity and meanwhile cut the communication overhead,
provided that the sender knows the secret quantum
information belongs to some special ensembles.\\

\noindent {\bf 2 Two tripartite QIS schemes}

Let the three legitimate parties be Alice, Bob and Charlie. Alice is
the initial owner and sender of a qubit state which can be written
as
\begin{eqnarray}
|\xi\rangle=\alpha'|0\rangle+\beta|1\rangle,
\end{eqnarray}
where, without loss of generality, $\alpha'$ is taken to be
real and $\beta$ complex, such that
$|\alpha'|^2+|\beta|^2=1$. Both $\alpha'$ and $\beta$ are
known to Alice, but not to Bob and Charlie.  $|\xi>$ can be
represented as a point on the Bloch sphere, which is
specified by two real angular coordinates $\theta$ and
$\phi$, such that $\alpha'=\cos (\theta /2)$ and
$\beta=\sin (\theta /2) e^{i \phi}$. If $|\xi\rangle$ is
chosen to be on the equatorial circle, then $\theta=\pi/2$
and accordingly $\alpha'=1/\sqrt2$ and $\beta= e^{i
\phi}/\sqrt2$ [16]; it is called an equatorial state. On
the other hand, if the azimuthal angle $\phi=0$ or $\pi$,
then the parameters $\alpha'$ and $\beta$ are both real; it
is called a real state. Now Alice wants to split the
quantum information/state into two shares and distribute
them to the receivers Bob and Charlie, such that the
original information can be deterministically recovered
only when the two receivers collaborate. Separately, each
individual receiver should have no knowledge of the secret.
In the following we will present two new schemes to achieve
this goal. The quantum channel linking the three parties
will be taken to be a GHZ state and asymmetric W state
respectively.
\\

\noindent {\bf 2.1 QIS with a shared GHZ state}

In this subsection we present a QIS scheme when the shared
quantum channel is a GHZ state. For short, this scheme will
be referred to as ZC1 scheme hereafter. In the beginning,
Alice, Bob, and Charlie own respectively qubit $a$, $b$,
and $c$, which are in a GHZ state:
\begin{eqnarray}
|\Psi\rangle_{abc}=\frac{1}{\sqrt{2}}\,
\Big(\,|000\rangle_{abc}+|111\rangle_{abc}\Big).
\end{eqnarray}
Since Alice knows her quantum information exactly, she can
perform a single-qubit measurement on her qubit $a$ in the
basis $\{|\xi\rangle,|\xi_{\perp}\rangle\}$, where
$|\xi_{\perp}\rangle=\beta^{*}|0\rangle-\alpha'|1\rangle$.
$|\Psi\rangle_{abc}$ can be rewritten as
\begin{eqnarray}
|\Psi\rangle_{abc}&=&\frac{1}{\sqrt
2}\,\Big[\,|\xi\rangle_{a}\big(\alpha'
|00\rangle_{bc}+\beta^* |11\rangle_{bc}\big)+
|\xi_{\perp}\rangle_{a}\big(\beta |00\rangle_{bc}-\alpha'
|11\rangle_{bc}\big)\,\Big].
\end{eqnarray}
If she gets $|\xi\rangle_a$, she publishes a single cbit
``0"; otherwise, she publishes ``1". This classical bit
informs Bob and Charlie the collapsed state of their two
qubits after Alice's measurement. 0 corresponds to the
state $\alpha' |00\rangle_{bc}+\beta^* |11\rangle_{bc}$
while 1 to  $\beta |00\rangle_{bc}-\alpha'
|11\rangle_{bc}$.

Without loss of generality, we will assume that the secret
quantum information is to be reconstructed with qubit $c$
at Charlie's site. In the case that Bob and Charlie get the
state $\beta |00\rangle_{bc}-\alpha' |11\rangle_{bc}$ (the
probability is 1/2), then they can collaborate to recover
the quantum information as follows. Bob performs a
single-qubit measurement in the $X$ basis $|\!\pm x\rangle
=\frac{1}{\sqrt{2}}\big(|0\rangle\pm|1\rangle\big)$ on his
qubit $b$ and informs Charlie of his result. Since,
\begin{eqnarray}
\beta |00\rangle_{bc}-\alpha'
|11\rangle_{bc}=\frac{1}{\sqrt 2}\,
\Big[\,|+x\rangle_b\big(\sigma_x^c\sigma_z^c\big)^\dag|\xi\rangle_{c}+
|-x\rangle_b\big(\sigma_x^c\big)^\dag|\xi\rangle_{c}\Big],
\end{eqnarray}
where $\sigma_x\equiv|1\rangle\langle 0|-|0\rangle\langle
1|$ and $\sigma_z\equiv |0\rangle\langle
0|-|1\rangle\langle 1|$ are Pauli operators, it is easy to
see that knowing Bob's result, Charlie can recover the
quantum information by performing an appropriate unitary
operation on his qubit $c$. Explicitly, corresponding to
Bob's outcomes $|+x\rangle_b$ and $|-x\rangle_b$, Charlie
performs the unitary operation $\sigma_x^c\sigma_z^c$ and
$\sigma_x^c$ respectively on his qubit $c$. In this case,
the quantum information is successfully shared and
reconstructed.

Now if Bob and Charlie get the state
$\alpha'|00\rangle_{bc}+\beta^* |11\rangle_{bc}$ instead
(again with probability 1/2), then in general they can not
recover the quantum information faithfully by any means.
However, if the secret quantum information is known to be
chosen from a special ensemble of equatorial or real qubits
states, then reconstruction becomes possible. (i) As
mentioned before, $\theta=\pi/2$ for equatorial states,
such that $|\xi\rangle=(|0\rangle+ e^{i
\phi}|1\rangle)/\sqrt2$ (i.e., $\alpha'=1/\sqrt2$ and
$\beta= e^{i \phi}/\sqrt2$). In this case, the state
$\alpha'|00\rangle_{bc}+\beta^* |11\rangle_{bc}$ becomes
$(|00\rangle_{bc}+ e^{-i \phi}|11\rangle_{bc})/\sqrt2$. Bob
and Charlie can recover the quantum information from this
state as follows. Bob carries out a single-qubit
measurement in the $X$ basis on his qubit $b$ and tells
Charlie his result. Since
\begin{eqnarray}
|00\rangle_{bc}+ e^{-i \phi}|11\rangle_{bc} =e^{-i\phi}\,
\Big[\,|+x\rangle_b \big(\sigma_x^c\sigma_z^c\big)^\dag
|\xi\rangle_{c}+
|-x\rangle_b\big(\sigma_x^c\big)^\dag|\xi\rangle_{c}\Big],
\end{eqnarray}
it is then clear that knowing Bob's result, Charlie can
recover the quantum information except for an unimportant
overall phase factor by performing an appropriate unitary
operation. Explicitly, corresponding to Bob's measurement
results $|+x\rangle_b$ and $|-x\rangle_b$, Charlie performs
respectively the unitary operation $\sigma_x\sigma_z$ and
$\sigma_x^{c}$ on qubit $c$ in his possession. (ii)When
Alice's secret quantum information is a real qubit state
(i.e., both $\alpha'$ and $\beta$ are real), then
$\alpha'|00\rangle_{bc}+\beta^* |11\rangle_{bc}$ becomes
$\alpha'|00\rangle_{bc}+\beta |11\rangle_{bc}$, which is in
essence the redundant state of the original quantum
information. Since
\begin{eqnarray}
\alpha'|00\rangle_{bc}+\beta |11\rangle_{bc}
=\frac{1}{\sqrt 2}\,\Big(|+x\rangle_b|\xi\rangle_{c}+
|-x\rangle_b\big(\sigma_z^c\big)^\dag|\xi\rangle_{c}\Big),
\end{eqnarray}
Bob and Charlie can again reconstruct Alice's secret quantum
information, using the procedure describe for case (i).

Hence, given a quantum channel of a GHZ state shared among
the three parties, and if the quantum information (known to
Alice) is chosen from a special ensemble of equatorial or
real states, then our new QIS protocol requires only a
single-qubit measurement and one cbit of classical
communication.

Lastly we examine the case where the shared three-qubit
state is not maximally entangled. Let the shared state
$|\Psi'\rangle_{abc}$ be given by
 \begin{equation}
 |\Psi'\rangle_{abc}=
 a|000\rangle_{abc}+b|111\rangle_{abc},
 \end{equation}
where $a$ and $b$ are real (if not, they can be made real
by a simple unitary rotation) and $a\ne b$. Intuitively, as
for teleportation, one would expect that in this case Bob
and Charlie can still recover the secret qubit, but with
probability less than unity. Indeed, following [42], it is
straightforward to show that the recovery probability is $2
a^2 b^2$.
\\

\noindent {\bf 2.2 QIS with a shared asymmetric W state}

Here we present another new QIS scheme when the shared
quantum channel is an asymmetric W state. It will be
referred to as the ZC2 scheme hereafter. As before, Alice,
Bob and Charlie respectively own qubits $a$, $b$, and $c$;
and the three qubits are in an asymmetric W state given by
\begin{eqnarray}
|\Phi\rangle_{abc}=\frac{1}{2}|001\rangle_{abc}+
\frac{1}{2}|010\rangle_{abc}+\frac{1}{\sqrt{2}}|100\rangle_{abc}.
\end{eqnarray}
It is easy to show that this state can be rewritten as
\begin{eqnarray}
|\Phi\rangle_{abc}=\frac{1}{\sqrt 2}\,
|\xi\rangle_{a}\Big[\frac{\alpha'}{\sqrt 2}\,
\big(|01\rangle_{bc}+|10\rangle_{bc}\big)+ \beta^*
|00\rangle_{bc}\Big]-\frac{1}{\sqrt 2}\,
|\xi_{\perp}\rangle_{a} \Omega_{bc}
\Big[|0\rangle_{b}(\sigma_{z}^c) |\xi\rangle_{c}\big],
\end{eqnarray}
where
\begin{eqnarray}
\Omega=|00\rangle\langle 00|+|11\rangle\langle 11| +
\frac{1}{\sqrt 2} \Big(|01\rangle\langle 01|
+|10\rangle\langle 01| +|01\rangle\langle 10| -
|10\rangle\langle 10| \Big).
\end{eqnarray}
From this equation, we see that, if Alice performs a
single-qubit measurement on qubit $a$ in the basis
$\{|\xi\rangle,|\xi_{\perp}\rangle\}$, then the state of
qubits $b$ and $c$ will collapse to either
$\frac{\alpha'}{\sqrt 2} (|01\rangle_{bc}+|10\rangle_{bc})+
\beta^* |00\rangle_{bc})$ or  $\Omega_{bc}^\dag
[|0\rangle_{b}(\sigma_{z}^c)^\dag |\xi\rangle_{c}]$. Bob
and Charlie know what the state is if Alice announces her
measurement result by broadcasting a one-cbit message as
described earlier.  In the case of $\Omega_{bc}^\dag
[|0\rangle_{b}(\sigma_{z}^c)^\dag |\xi\rangle_{c}]$, Bob
and Charlie can collaborate and recover the quantum
information by first performing the unitary operation
$\Omega$ on their qubits $b$ and $c$. Subsequently Charlie
applies the unitary operation $\sigma_{z}^\dag$ on his
qubit $c$ and obtains $|\xi\rangle_c$. On the other hand,
if the collapsed state is $\frac{\alpha'}{\sqrt 2}
(|01\rangle_{bc}+|10\rangle_{bc})+ \beta^*
|00\rangle_{bc})$, then in general, Bob and Charlie cannot
work together to recover the quantum information, because
the general complex coefficient $\beta^*$ can not be
converted into $\beta$ via an unitary operations. However,
as before, if the secret quantum information is chosen from
a special ensemble of equatorial or real states, then
recovery again becomes possible. (I) If $|\xi\rangle$ is an
equatorial state, i.e., $\alpha'=1/\sqrt2$ and $\beta= e^{i
\phi}/\sqrt2$, then $\frac{\alpha'}{\sqrt 2}
(|01\rangle_{bc}+|10\rangle_{bc})+ \beta^* |00\rangle_{bc}$
becomes $\frac{1}{2} (|01\rangle_{bc}+|10\rangle_{bc})+
\frac{e^{-i \phi}}{\sqrt 2} |00\rangle_{bc}$. Bob and
Charlie can reconstruct $|\xi\rangle$ at Charlie's site by
performing the joint unitary operation $\Omega$ on their
qubits $b$ and $c$. The resultant state is given by
$\frac{1}{\sqrt 2}e^{-i \phi}|0\rangle_{b}(e^{i
\phi}|1\rangle_{c}+ |0\rangle_{c})=\frac{1}{\sqrt 2}e^{-i
\phi}|0\rangle_{b}|\xi\rangle_c$, indicating that Charlie
now holds the secret quantum information in his hand. (II)
If $|\xi\rangle$ is a real state, i.e., both $\alpha'$ and
$\beta$ are real, then $\frac{\alpha'}{\sqrt 2}
(|01\rangle_{bc}+|10\rangle_{bc})+ \beta^* |00\rangle_{bc}$
becomes $\frac{\alpha'}{\sqrt 2}
(|01\rangle_{bc}+|10\rangle_{bc})+ \beta |00\rangle_{bc}$.
Bob and Charlie can recover the the quantum information by
first performing the unitary operation $\Omega$ on their
qubits $b$ and $c$ together. Then Charlie can recover the
quantum information by applying the unitary operation
$\sigma_{x}\sigma_{z}$ on his qubit $c$.

Hence again, we have shown that, to split a qubit chosen
from a special ensemble (equatorial or real) and known to
Alice, she only needs to perform a single-qubit measurement
and publish one cbit, provided that the three parties share
an asymmetric tripartite W state.\\

Finally we examine the case where the shared three-qubit
state is not maximally entangled. Following the above
procedure, we found that if the shared state
$|\Phi'\rangle_{abc}$ is of the form given by
 \begin{equation}
 |\Phi'\rangle_{abc}=
 a |001\rangle_{abc}+b |010\rangle_{abc}
 +c |100\rangle_{abc},
 \end{equation}
where $a\ne b\ne c$, then it is in general not possible for
Bob and Charlie to recover the original secret qubit. In
the special case of $a=b\ne c$, our procedure works if
$c=\sqrt{2}a$, which however just brings
$|\Phi'\rangle_{abc}$ back to the original asymmetric W
state $|\Phi\rangle_{abc}$.

\noindent {\bf 3 Discussions}

In the last section we have proposed two QIS schemes, ZC1 and ZC2,
using respectively a tripartite GHZ state and an asymmetric W state
as the quantum channel. Here we compare them with two other QIS
schemes, the HBB scheme[28] and the Zheng scheme[39], using
respectively the same quantum channels as ZC1 and ZC2.

As we shall see, in ordinary QIS schemes like those of HBB
and Zheng, Alice must make a Bell state measurement and
send two cbits to Bob and Charlie. In the schemes we have
proposed, however, Alice needs only to make a single-qubit
measurement and announce a one-bit classical information.
With these simplifications, our schemes are subjected to
two restrictions: (1) Alice must know the qubit being sent,
and (2) the secret qubit must be a member of a special
ensemble which is public knowledge. In most practical
situations, the sender Alice is the owner of the secret
qubit, so the first restriction is not a trade-off in most
cases. With regard to the second restriction, the secret
information in our schemes must be in the form of either an
equatorial or real qubit, which is characterized by a
single parameter, $\phi$ or $\theta$. In contrast, ordinary
QIS schemes work for an arbitrary secret qubit
characterized by two parameters $\phi$ and $\theta$. This
loss of freedom is a trade-off that we have to make in our
schemes.

Finally, we also discuss the case of Alice withholding the
classical information as a control [43]. \\

\noindent {\bf 3.1 Comparison of ZC1 scheme with HBB scheme }

In 1999, Hillery, B\v{u}zek and Berthiaume[28] first presented a QIS
scheme (HBB scheme) using GHZ states as the quantum channel. In the
HBB QIS scheme, Alice's secret quantum information to be shared
between Bob and Charlie is given by
\begin{eqnarray}
|u\rangle_x=\alpha|0\rangle_x+\beta|1\rangle_x,
\end{eqnarray}
where $\alpha$ and $\beta$ are complex and satisfy
$|\alpha|^2+|\beta|^2=1$. In addition, Alice owns another
qubit $a$ which forms a GHZ state with Bob's qubit $b$ and
Charlie's qubit $c$. The combined four-qubit state is given
by
\begin{eqnarray}
|\Lambda\rangle_{xabc}=|u\rangle_x\otimes|\Psi\rangle_{abc},
\end{eqnarray}
which can be rewritten as
\begin{eqnarray}
|\Lambda\rangle_{xabc}&=&\frac{1}{2}\,\Big[\,|\psi^+\rangle_{xa}
\big(\alpha |00\rangle_{bc}+\beta |11\rangle_{bc}\big)+
|\psi^-\rangle_{xa}\big(\alpha |00\rangle_{bc}-\beta
|11\rangle_{bc}\big)
\nonumber \\
&&+|\phi^+\rangle_{xa}\big(\beta |00\rangle_{bc}+\alpha
|11\rangle_{bc}\big)+ |\phi^-\rangle_{xa}\big(\beta
|00\rangle_{34}-\alpha |11\rangle_{bc}\big)\Big],\nonumber \\
&=&\frac{1}{2\sqrt{2}}\,\Big[\,|\psi^+\rangle_{xa}|+
x\rangle_b|u\rangle_c + |\psi^+\rangle_{xa}|-x\rangle_b
\big(\sigma_z^c\big)^\dag|u\rangle_c \nonumber \\
&&+|\psi^-\rangle_{xa}|+x\rangle_b
\big(\sigma_z^c\big)^\dag|u\rangle_c +
|\psi^-\rangle_{xa}|-x\rangle_b |u\rangle_c \nonumber\\
&&+|\phi^+\rangle_{xa}|+x\rangle_b
\big(\sigma_x^c\sigma_z^c\big)^\dag|u\rangle_c
+|\phi^+\rangle_{xa}|-x\rangle_b
\big(\sigma_x^c\big)^\dag|u\rangle_c \nonumber\\ &&-
|\phi^-\rangle_{xa}|+x\rangle_b \big(\sigma_x^c\big)^\dag
|u\rangle_c- |\phi^-\rangle_{xa}|-x\rangle_b
\big(\sigma_x^c\sigma_z^c\big)^\dag |u\rangle_c\Big],
\end{eqnarray}
where $|\psi^{\pm}\rangle$ and $|\phi^{\pm}\rangle$ are the
four Bell states given by
\begin{eqnarray}
|\psi^{\pm}\rangle=\frac{1}{\sqrt{2}}\big(|00\rangle \pm
|11\rangle\big), \ \
|\phi^{\pm}\rangle=\frac{1}{\sqrt{2}}\big(|01\rangle \pm
|10\rangle\big).
\end{eqnarray}
Therefore Alice can split the quantum information and have
it shared by Bob and Charlie by first performing a
two-qubit Bell measurement on qubits $x$ and $a$, and then
announcing the outcome as a two cbits message. Knowing
Alice's outcome, Bob and Charlie can easily recover the
quantum information via a similar procedure as already
described.

Note that, since the state $|u\rangle_x$ in Eq. (12) is
arbitrary, the HBB scheme can be used for splitting any
single-qubit quantum information including the equatorial
and real states. However, Independent of the nature of the
quantum information to be shared, the HBB requires the
sender Alice to perform a Bell measurement and publish two
cbits. It is clear that, in the case of equatorial or real
states, this is a waste of resources. As we have
demonstrated in subsection 2.1, given the same GHZ channel,
an equatorial or real qubit can be split and shared using
only a single-qubit measurement and one cbit of classical
communication. Hence, for these special qubit states, our
ZC1 scheme not only decreases the measurement complexity,
it also reduces the required classical communication by one
half.
\\

\noindent {\bf 3.2 Comparison of ZC2 scheme with Zheng
scheme}

In 2006, Zheng[39] presented a QIS scheme (Zheng scheme)
using an asymmetric W state as shown in the Eq. (8). This
scheme also works for arbitrary secret quantum information
as shown in Eq. (12). In this case, the joint state of
Alice's two qubits $x$ and $a$, Bob's qubit $b$ and
Charlie's qubit $c$ is given by
\begin{eqnarray}
\Gamma_{xabc}&=&|u\rangle_x|\Phi\rangle_{abc}\nonumber\\
&=&\big(\alpha|0\rangle_x+\beta|1\rangle_x\big)
\Big(\frac{1}{2}|001\rangle_{abc}+\frac{1}{2}|010\rangle_{abc}+
\frac{1}{\sqrt{2}}|100\rangle_{abc}\Big).
\end{eqnarray}
It can also be rewritten as
\begin{eqnarray}
\Gamma_{xabc}&=& \frac12\,
|\psi^+\rangle_{xa}\,\Omega_{bc}\Big[\,|0\rangle_{b}
\big(\sigma_x^c\sigma_z^c\big)^\dag|u\rangle_c\Big]+
 \frac12\,
|\psi^-\rangle_{xa}\,\Omega_{bc}\,\Big[\,|0\rangle_{b}
\big(\sigma_x^c\big)^\dag|u\rangle_c\Big]\nonumber\\
&&+ \frac12\,
|\phi^+\rangle_{xa}\,\Omega_{bc}\,\Big[\,|0\rangle_{b}
|u\rangle_c\Big]+ \frac12\,
|\phi^+\rangle_{xa}\,\Omega_{bc}\Big[\,|0\rangle_{b}
\big(\sigma_z^c\big)^\dag|u\rangle_c\Big].
\end{eqnarray}
Hence again Alice can split the quantum information and
distribute the shares to Bob and Charlie by performing a
two-qubit Bell measurement and publishing the outcome in
the form of a two-cbit message. With this message, Bob and
Charlie can then collaborate and reconstruct the quantum
information as described previously.

As in the HBB scheme, this scheme works for arbitrary qubit
states, and it requires the sender Alice to perform a Bell
state measurement and broadcast the outcome as a two-cbit
message. In comparison, our new scheme ZC2 works for
equatorial and real states only. This restriction is
however compensated by the fact that Alice needs to make
only a single qubit measurement, and broadcast a one-cbit
message. Hence, in cases where the secret state is chosen
from the ensemble of equatorial or real states, and it is
known to Alice, the ZC2 scheme is more preferable than the
Zheng scheme.\\

\noindent {\bf 3.3 Controlled quantum information
splitting}

In ZC1, if the secret quantum information is an equatorial
state, then the secret resides in the phase $\phi$.  After
Alice announces her one-cbit message, Bob or Charlie has
absolutely no information of the secret if they do not
collaborate.  This can be seen from the fact that the
density matrices of the qubits in their hands are both
$I/2$ which is independent of $\phi$.  In all other cases
discussed, the density matrices of Bob's and Charlie's
qubits are not $I/2$. That means each of them has some
information of the secret even if they do not collaborate.
This is undesirable in a QIS scheme.

This drawback can be avoided if Alice delays her classical
communication.  A similar scenario was first proposed by
Cheung in 2006[43]. The point is, in ordinary QIS schemes,
the sender of the secret has no control over the use of the
secret state after it is distributed. That is, the
receivers can come together and reconstructed it anytime
they want. In a controlled quantum information splitting
(or secret sharing) scheme, by withholding the classical
communication, the sender (or controller) can decide when
the secret information is to be recovered by the receivers.
Before that, the receivers has absolutely no knowledge of
the secret information, and it is impossible for them to
recover the secret even if they want to.

Therefore, except for the case of ZC1 with equatorial
qubits, Alice should not announce the one-cbit message
right after her measurement.  Instead she should withhold
the information until the time when she wants Bob and
Charlie to reconstructed the secret state.  Then it is
clear that Bob and Charlie separately knows nothing about
the secret before they start the reconstruction process. It
is important to note that here the delay of the classical
communication is not merely an option, but is necessary to
protect the secret information from leaking to Bob and
Charlie separately.\\

\noindent {\bf 4 Summary and Conclusion}

To summarize, we have proposed two QIS schemes (ZC1 and
ZC2) using respectively GHZ and asymmetric W states as
quantum channels. The scheme works for qubits chosen from a
special ensemble of equatorial or real states, moreover the
state should be known to the sender Alice. In these new
schemes, Alice is only required to perform a single qubit
measurement and broadcast a one-cbit message. Clearly no
other scheme can possibly achieve the same goal with
simpler measurement and less classical communication. In
the existing QIS schemes (HBB[28] and Zheng[39]) using the
same quantum channels, Alice must make a two-qubit joint
measurement and announce a two-cbit message. In return for
the more complex measurement and greater amount of
classical communication, these schemes work for arbitrary
quantum states which need not be known to the sender Alice.
However it is reasonable to assume that, in most practical
situations, Alice is the owner of the secret and so she
knows what the secret is. Furthermore if the secret state
belongs to the special ensemble of equatorial or real
states, then our schemes are clearly more desirable.

With regard to experimental implementations, the main
difficulty lies in the establishment of a three-qubit
entangled state among three remote sites. While GHZ states
have been observed in laboratories [44,45,46], creating and
maintaining remote entanglements is still something to be
desired. This obstacle is however common to our schemes as
well as others. Nevertheless, as we saw, our schemes
requires only single-qubit measurements which, with present
date technologies, are immensely simpler and more efficient
to perform than two-qubits ones. In this regard, our
schemes are relatively much easier to realize
experimentally. Note also that, in order to be able to
recover the secret qubit, in general Bob and Charlie must
keep their qubits entangled after Alice's measurement. This
is however not necessarily if the initial state is a GHZ
state.  In this case, since Bob needs to make a measurement
in the $\{|\pm x\rangle\}$ basis independent of Alice's
outcomes, he could do it anytime but announces his result
only when he wants Charlie to recovered the secret qubit.
Then there is no need to maintain the entanglement between
Bob and Charlie for an indefinite length of time.

Finally we discussed the case of Alice withholding the
classical information as a control[43]. For ZC1 with
equatorial states, this is an option which allows Alice to
decide if and when Bob and Charlie should start the
reconstruction process. For all other cases, however,
delaying the announcement of the classical information is
also necessary to ensure that, before Bob and Charlie start
to work together to reconstruct the secret qubit, each of
them has no knowledge of the secret whatsoever.\\

\noindent {\bf Acknowledgements}

This work is partly supported by the program for New
Century Excellent Talents at the University of China under
Grant No.NCET-06-0554, the National Natural Science
Foundation of China under Grant No.60677001, the
science-technology fund of Anhui province for outstanding
youth under Grant No.06042087, the key fund of the ministry
of education of China under Grant No.206063, and the
general fund of the educational committee of Anhui province
under Grant No.2006KJ260B.\\

\noindent {\bf References}

\noindent[1] H. K. Lo,  Phys. Rev. A {\bf62} (2000) 012313.

\noindent[2] J. Eisert, K. Jacobs, P. Papadopoulos, and M. B.
Plenio, Phys. Rev. A {\bf 62} (2000) 052317.

\noindent[3] D. Collins, N. Linden, and S. Popescu, Phys. Rev. A
{\bf 64} (2001) 032302.

\noindent[4] C. H. Bennett, G. Brassard, C. Crepeau,  R. Jozsa, A.
Peres, and W. K. Wotters, Phys. Rev. Lett. {\bf70} (1993) 1895.

\noindent[5] M. A. Nielsen and I. L. Chuang, Quantum Computation and
Quantum Information (Cambridge University Press, Cambridge, 2000).

\noindent[6] M. S. Zubairy, Phys. Rev. A  {\bf58} (1998) 4368.

\noindent[7] S. Stenholm and P. J. Bardroff, Phys. Rev. A  {\bf58}
(1998) 4373.

\noindent[8] J. Lee and M. S. Kim, Phys. Rev. Lett.  {\bf84} (2000)
4236.

\noindent[9] P. Agrawal, A. K. Pati, Phys. Lett. A  {\bf305} (2002)
12 .

\noindent[10] L. Roa, A. Delgado, and I. Fuentes-Guridi, Phys. Rev.
A {\bf68} (2003) 022310.

\noindent[11] G. Rigolin, Phys. Rev. A {\bf71} (2005) 032303.

\noindent[12] Z. J. Zhang and Z. X. Man, Phys. Lett. A {\bf341}, 55
(2005); Z. J. Zhang, Phys. Lett. A  {\bf351} (2006) 55 .

\noindent[13] Y. Yeo and W. K. Chua, Phys. Rev. Lett. {\bf96} (2006)
060502; Y. Yeo, Phys. Rev. A  {\bf74} (2006) 052305.

\noindent[14] P. X. Chen, S. Y. Zhu, and G. C. Guo, Phys. Rev. A
{\bf74} (2006) 032324.

\noindent[15] Z. J. Zhang, Y. M. Liu, and D. Wang, Phys. Lett. A
(doi:10.1016/j.physleta.2007.07.017).

\noindent[16] A. K. Pati, Phys. Rev. A {\bf63} (2001) 014302.

\noindent[17] C. H. Bennett, D. P. DiVincenzo, P. W. Shor, J. A.
Smolin, B. M. Terhal, and W. K. Wootters, Phys. Rev. Lett {\bf87}
(2001) 077902.

\noindent[18] I. Devetak and T. Berger, Phys. Rev. Lett {\bf87}
(2001) 177901.

\noindent[19] D. W. Berry and B. C. Sanders, Phys. Rev. Lett {\bf90}
(2003) 027901.

\noindent[20] D. W. Leung, P. W. Shor, Phys. Rev. Lett {\bf90}
(2003) 127905.

\noindent[21] A. Hayashi, T. Hashimoto, and M. Horibe, Phys. Rev. A
{\bf67} (2003) 052302.

\noindent[22] A. Abeyesinghe and P. Hayden, Phys. Rev. A {\bf68}
(2003) 062319.

\noindent[23] M. Y. Ye, Y. S. Zhang, and G. C. Guo, Phys. Rev. A
{\bf69} (2004) 022310.

\noindent[24] Z. Kurucz, P. Adam, Z. Kis, and J. Janszky, Phys. Rev.
A {\bf72} (2005) 052315.

\noindent[25] G. Y. Xiang, J. Li, Y. Bo, and G. C. Guo, Phys. Rev. A
{\bf72} (2005) 012315.

\noindent[26] N. A. Peters, J. T. Barreiro, M. E. Goggin, T. C. Wei,
and P. G. Kwiat, Phys. Rev. Lett {\bf94} (2005) 150502.

\noindent[27] Z. Kurucz, P. Adam, and J. Janszky, Phys. Rev. A
{\bf73} (2006) 062301.

\noindent[28] M. Hillery, V. B\v{u}zek, and A. Berthiaume, Phys.
Rev. A {\bf59} (1999) 1829.

\noindent[29] A. Karlsson, M. Koashi, and N. Imoto, Phys. Rev. A
{\bf 59} (1999) 162 .

\noindent[30] R. Cleve, D. Gottesman, H.K. Lo, Phys. Rev. Lett. {\bf
83} (1999) 648.

\noindent[31] D. Gottesman, Phys. Rev. A {\bf 61} 042311 (2000).
042311.

\noindent[32] S. Bandyopadhyay, Phys. Rev. A {\bf 62}, (2000)
012308.

\noindent[33] L. Y. Hsu, Phys. Rev. A {\bf 68} (2003) 022306; L. Y.
Hsu, C. M. Li, Phys. Rev. A {\bf 71} (2005) 022321.

\noindent[34] A. M. Lance et al, Phys. Rev. Lett. {\bf 92} (2004)
177903; A. M. Lance et al, Phys. Rev. A {\bf 71} (2005) 033814.

\noindent[35] Z. J. Zhang et al., Phys. Rev. A {\bf 71} (2005)
044301; Z. J. Zhang et al., Eur. Phys. J. D {\bf 33} (2005) 133.

\noindent[36] S. K. Singh and R. Srikanth, Phys. Rev. A {\bf 71}
(2005) 012328.

\noindent[37] F. G. Deng, X. H. Li, C. Y. Li, and H. Y. Zhou, Phys.
Rev. A {\bf 72} (2005) 044301; F. G. Deng, X. H. Li, C. Y. Li, P.
Zhou, and H. Y. Zhou, Eur. Phys. J. D {\bf 39} (2006) 459.

\noindent[38] G. Gordon and G. Rigolin, Phys. Rev. A {\bf 73} (2006)
062316.

\noindent[39] S. B. Zheng, Phys. Rev. A {\bf 74} (2006) 054303.

\noindent[40] Z. Y. Wang, Y. M. Liu, and Z. J. Zhang,  Opt.
Commun. {\bf 276} (2007) 322.

\noindent[41] Z. Y. Wang, H. Yuan, S. H. Shi, and Z. J.
Zhang, Eur. Phys. J. D {\bf 41} (2007) 371.

\noindent[42] B.-S. Shi, Y.-K. Jiang, G.-C. Guo, Phys.
Lett. A, \textbf{268} (2000), 161.

\noindent[43] C. Y. Cheung, Phys. Scr. {\bf 74} (2006) 459.

\noindent[44] D. Leibfried et al., Nature {\bf 438} (2005)
639.

\noindent[45] J.-W. Pan et al, Nature {\bf 403} (2000) 515.

\noindent[46] A. Rauschenbeutel et al., Science {\bf 288}
(2000)2024.

\enddocument